\documentclass[aps,prd,preprint,nofootinbib]{revtex4}
\usepackage{epsfig}
\usepackage{graphicx}

\begin{document}
\title{Hadronic Production of $B_c(B^*_c)$ Meson
Induced by the Heavy Quarks inside the Collision Hadrons\\[6mm]}
\author{Chao-Hsi Chang$^{1,2}$ \footnote{email:
zhangzx@itp.ac.cn}, Cong-Feng Qiao$^{1,3}$\footnote{email:
qiaocf@gucas.ac.cn}, Jian-Xiong Wang$^{4}$\footnote{email:
jxwang@mail.ihep.ac.cn}, Xing-Gang Wu$^{2,4}$\footnote{email:
wuxg@itp.ac.cn}}
\address{$^1$CCAST (World Laboratory), P.O.Box 8730, Beijing 100080,
P.R.China.\\
$^2$Institute of Theoretical Physics, Chinese Academy of Sciences,
P.O.Box 2735, Beijing 100080, P.R.China.\\
$^3$Department of Physics, Graduate School of the Chinese\\
Academy of Sciences, Beijing 100049, P.R.China\\
$^4$Institute of High Energy Physics, P.O.Box 918(4), Beijing
100049, P.R.China}

\begin{abstract}
In general-mass variable-flavor-number (GM-VFN) scheme, the `heavy
quark mechanisms' of $B_c(B_c^*)$ meson hadroproduction via the
sub-processes $g+c\to B_c(B_c^*)+\cdots$ and $g+\bar{b}\to
B_c(B_c^*)+\cdots$ are investigated. In the scheme, possible
double counting from these mechanisms and the `light mechanisms'
(the gluon-gluon fusion and quark-antiquark annihilation
mechanisms) is subtracted properly. The numerical results show
that the transverse momentum, $p_t$, distribution of the produced
$B_c(B_c^*)$ from the `intrinsic heavy quark mechanisms' (i.e. the
heavy quark mechanisms in which the double counting components
have been subtracted accordingly) declines faster with the
increment of the $p_t$ than that from the `light mechanisms', and
only in small $p_t$ region ($p_t\lesssim 7.0$ GeV) the `intrinsic
heavy quark mechanisms' themselves may make remarkable
contributions. The combined contributions from the `intrinsic
heavy quark mechanisms' and the `light mechanisms' to the
production are compared with that obtained by the most
calculations in literature, which in some sense are within the
fixed flavor number (FFN) scheme at leading order, and we find
that the production by virtue of the GM-VFN scheme is more or
less the same as the one in literature, except in the small
$p_t$ region.\\

\noindent {\bf PACS numbers:} 12.38.Bx, 13.85.Ni, 14.40.Nd,
14.40.Lb.

\noindent {\bf Keywords:} inclusive hadroproduction, intrinsic
mechanism, $B_c$ meson.

\end{abstract}

\maketitle

\section{introduction}

The unique and stable `double heavy-flavored' meson, $B_c$, has
been confirmed in experiment \cite{CDF,CDF1,D0}, and the
observations are consistent with the theoretical expectations
within theoretical uncertainties and experimental errors. In view
of the prospects of $B_c$ production at Fermilab Tevatron (Run-II)
and at LHC, the $B_c$ physics is compelling. The future copious
data require more accurate theoretical predictions, especially,
that on the production (at Tevatron and LHC). The hadronic
production of $B_c$ meson has been studied quite a lot
\cite{prod,prod0,prod1,prod2,prod3,prod4,changwu,prod5,prod6,bcvegpy1,bcvegpy2}
already. It is remarkable that with special techniques the event
generator BCVEGPY \cite{bcvegpy1,bcvegpy2} for the production is
in compliment to the PYTHIA environment \cite{pythia} and powerful
enough in generating event samples for most purposes, i.e., with
it one can enhance the event generating efficiency greatly in
contrast to PYTHIA itself. With newly upgraded version BCVEGPY2.0,
not only the ground state of $B_c$ meson but also `low laying'
excited states can be generated.

Up to now, the predictions for the $B_c$ hadronic production are
mainly based on the dominant gluon-gluon fusion mechanism in terms
of perturbative QCD (pQCD) calculations in order of $\alpha_s^4$,
i.e., via the sub-process $g+g\rightarrow (c\bar{b})+b+\bar{c}$
with $(c\bar{b})$ in the configurations of $S$- or $P$-wave
states. The less important mechanism via light quark-antiquark
annihilation $q+\bar{q}\rightarrow (c\bar{b})+b+\bar{c}$ is
studied in \cite{prod1,changwu} only for comparison. However, the
approaches to the existent estimates, in fact, all are in the
fixed flavor number (FFN) scheme \cite{ffn1,ffn2,ffn3} only with
certain extension\footnote{As for the exact FFN scheme for
$b$-production, the active quark flavor numbers in the initial
state is limited to $N_f=3$ (i.e. the initial state does not
contain the heavy quark components at all) and accordingly in the
production of $B_c(B_c^*)$, the PDFs, just CTEQ5F3\cite{cteq5},
which corresponds to the 3-flavor scheme of FFN, should be used;
but in the existent estimates when extending to higher anergy and
higher $p_t$ for the production, CTEQ6L (or CTEQ4L) etc
\cite{6lcteq}, which is derived with $N_f>3$ as the PDFs, was used
instead, thus here we call such approach as extended FFN scheme.}.
Recently, as pointed out by the authors of Ref. \cite{qiao}, in
hadronic production of charmonium, the contribution from the sea
charm quark via the subprocess $c+g\to (c\bar{c})+c$ may be
greater than that from the gluon-gluon fusion via the subprocess
$g+g\to (c\bar{c}) +\bar{c}+c$ for charmonium color-singlet
production. Therefore, it is interesting to examine the production
by applying the general-mass variable-flavor-number (GM-VFN)
scheme \cite{acot,gmvfn1,gmvfn2} and see how important are the
heavy charm and bottom quark mechanisms in the hadronic production
of $B_c(B_c^*)$ meson not only to compare with the gluon-gluon
fusion mechanism but also with the existent estimations. To be
noted that here the heavy quark means the sea parton for the heavy
flavor, which is different from what it means in Ref.
\cite{brodsky}. Generally the mechanisms induced by the heavy
quark components must be considered in the GM-VFN scheme, and they
are suppressed in comparison with the light partons such as the
light valance quarks, light sea quarks and gluons in PDFs.
However, the suppression fact may be `compensated' by their lower
order nature in perturbative QCD and a much `greater' phase space.
Naively, the sub-processes $c+g\to (c\bar{b})+b$ and $\bar{b}+g\to
(c\bar{b})+\bar{c}$ are $2$-body $\to 2$-body processes in the
order of ${\cal O}(\alpha_s^3)$, while the gluon-gluon fusion
subprocess  $g+g\rightarrow (c\bar{b})+b+\bar{c}$ is $2$-body $\to
3$-body process in the order of ${\cal O}(\alpha_s^4)$. Thus, we
devote this work to estimate the production in terms of the GM-VFN
scheme by taking into account the contributions from the `heavy
quark mechanisms' and that from the light partons, especially, the
gluon-gluon fusion mechanism together, not only estimating the
total cross sections, but also studying their properties on
transverse-momentum $p_t$ and rapidity $\eta$ distributions etc.

In GM-VFN scheme when we talk about the heavy quark components of
PDFs and taking into account both of the 'heavy quark mechanisms'
and the gluon-gluon fusion mechanism for the hadronic production,
one has to solve the double counting problem: i.e. a full QCD
evolved 'heavy quark' charm/bottom distribution functions,
according to the Altarelli-Parisi equations, includes all the
terms proportional to $\ln\left(\frac{\mu^2}{m^2_Q}\right)$ ($\mu$
the factorization scale and $m_Q$ the heavy quark mass); and some
of them come from the gluon-gluon fusion mechanism, i.e., a few
terms appear from the integration of the phase-space for the
gluon-gluon fusion mechanism. Therefore, one needs to make proper
subtractions to solve the double counting problem \cite{cwz}. One
convenient way to do the `subtraction' is to adopt the GM-VFN
scheme, in which the heavy-quark mass effects are treated in a
consistent way both for the hard scattering amplitude and the PDFs
\cite{acot,gmvfn1,gmvfn2}. Moreover, it will be necessary to use
the dedicated PDFs with heavy-mass effects included, which are
determined by global fitting utilizing massive hard-scattering
cross-sections. For instance, for the present analysis, the
CTEQ6HQ \cite{6hqcteq} is used. Later on for convenience, we will
call the `heavy quark mechanisms' which have been subtracted
according to method in GM-VFN scheme as `intrinsic ones'
accordingly.

The paper is organized as follows. In Sec.II, the basic formulae
for the `intrinsic charm and bottom mechanisms' in the GM-VFN
scheme are presented for the hadronic production of $S$-wave
$c\bar{b}$-quarkonium states, i.e. $B_c(^1S_0)$ and
$B^*_c(^3S_1)$. In Sec.III, we present the numerical results for
the 'intrinsic mechanisms' and make some comparisons about them.
The final section is reserved for discussion and summary.

\section{calculation technique}

To study the `intrinsic heavy quark mechanisms' for the hadronic
$B_c(B_c^*)$-production, we need to study the hadronic processes
of $g+c\to B_c+\cdots$, $g+\bar{b}\to B_c+\cdots$ and
$c+\bar{b}\to B_c+\cdots$. And, to make the 'intrinsic mechanisms'
and the gluon-gluon fusion mechanism consistently coexist, we need
to do the subtractions so as to avoid the double counting. For
this purpose we adopt the proper way in GM-VFN scheme
\cite{acot,gmvfn1,gmvfn2} in this work.

According to pQCD factorization theorem, in the GM-VFN scheme
\cite{acot,gmvfn1,gmvfn2}, the cross-section for the hadronic
production of $B_c(B^*_c)$ (including all of the mechanisms, even
the light-quark annihilation one via the subprocesses
$q_k+\bar{q_k}\to B_c+\bar{c}+b\;\;(k=u,d,s)$) is formulated as
\begin{eqnarray}
d\sigma&=&F^{g}_{H_{1}}(x_{1},\mu) F^{g}_{H_{2}}(x_{2},\mu)
\bigotimes d\hat{\sigma}_{gg\rightarrow
B_{c}(B_c^*)}(x_{1},x_{2},\mu)\nonumber\\
&+& \sum_{i,j=1,2(i\neq j),\;k}F^{q_k}_{H_{i}}(x_{1},\mu)
F^{\bar{q_k}}_{H_{j}}(x_{2},\mu)\bigotimes
d\hat{\sigma}_{q_k\bar{q_k}\rightarrow
B_{c}(B_c^*)}(x_{1},x_{2},\mu)\nonumber\\
&+& \sum_{i,j=1,2;i\neq
j}F^{g}_{H_{i}}(x_{1},\mu)\left[F^{c}_{H_{j}}(x_{2},\mu)-
F^{g}_{H_{j}}(x_{2},\mu)\bigotimes F^c_g(x_2,\mu)\right]
\bigotimes d\hat{\sigma}_{gc\rightarrow
B_{c}(B_c^*)}(x_{1},x_{2},\mu)
\nonumber\\
&+& \sum_{i,j=1,2;i\neq j}F^{g}_{H_{i}}(x_{1},\mu)
\left[F^{\bar{b}}_{H_{j}}(x_{2},\mu)-
F^{g}_{H_{j}}(x_{2},\mu)\bigotimes F^{\bar{b}}_g(x_2,\mu)\right]
\bigotimes d\hat{\sigma}_{g\bar{b}\rightarrow
B_{c}(B_c^*)}(x_{1},x_{2},\mu)\nonumber\\
&+& \sum_{i,j=1,2;i\neq j}\left[\left(F^{c}_{H_{i}}(x_{1},\mu) -
F^{g}_{H_{i}}(x_{1},\mu)\bigotimes F^{c}_g(x_1,\mu)\right)
\left(F^{\bar{b}}_{H_{j}}(x_{2},\mu)-
F^{g}_{H_{j}}(x_{2},\mu)\bigotimes
F^{\bar{b}}_g(x_2,\mu)\right)\right]\nonumber\\
&&\bigotimes d\hat{\sigma}_{c\bar{b}\rightarrow
B_{c}(B_c^*)}(x_{1},x_{2},\mu)+\cdots, \label{pqcdf0}
\end{eqnarray}
where the ellipsis means the terms in higher $\alpha_s$ order.
$F^{i}_{H}(x,\mu)$ (with $H=H_1$ or $H_2$ and $x=x_1$ or $x_2$) is
the distribution function of parton $i$ in hadron $H$. $d\sigma$
stands for the hadronic cross-section and $d\hat\sigma$ stands for
the corresponding subprocesses. For convenience, we have taken the
renormalization scale $\mu_R$ for the subprocess and the
factorization scale $\mu_F$ for factorizing the PDFs and the hard
subprocess to be the same, i.e. $\mu_R=\mu_F=\mu$. In the square
brackets, the subtraction term for $F^{Q}_{H}(x,\mu)$ is defined
as
\begin{equation}\label{subtraction}
F^{Q}_{H}(x,\mu)_{SUB}=F^{g}_{H}(x,\mu)\bigotimes
F^{Q}_g(x,\mu)=\int^1_{x}F^{Q}_g(\kappa,\mu)
F^{g}_{H}(\frac{x}{\kappa},\mu) \frac{d\kappa}{\kappa}.
\end{equation}
The quark distribution $F^Q_g(x,\mu)$ (with $Q$ stands for heavy
quark $c$ or $\bar{b}$) within an on-shell gluon up to order
$\alpha_s$ is connected to the familiar $g\to Q\bar{Q}$ splitting
function $P_{g\to Q}$, i.e.
\begin{equation}\label{quark}
F^Q_g(x,\mu)=\frac{\alpha_s(\mu)}{2\pi}\ln\frac{\mu^2}{m^2_Q}P_{g\to
Q}(x),
\end{equation}
with $P_{g\to Q}(x)=\frac{1}{2}(1-2x+2x^2)$.

In Eq.(\ref{pqcdf0}), the first term represents for the dominant
gluon-gluon fusion mechanism; the second one for the light quark
and anti-quark annihilation mechanism; the remainders for the so
`intrinsic charm/bottom mechanisms', in which the subtraction is
introduced to avoid the double counting \cite{acot}. The
gluon-gluon fusion mechanism and the light quark and anti-quark
annihilation mechanism have been considered in several previous
papers\cite{prod,prod0,prod1,prod2,prod3,prod4,prod5,prod6}, and
for their investigation, now the $B_c$ meson generator BCVEGPY
\cite{bcvegpy1,bcvegpy2} may be employed conveniently. As for the
`intrinsic charm/bottom mechanisms', we need to calculate three
kinds of subprocesses: $g+c\to B_c(B_c^*)+b$, $g+\bar{b}\to
B_c(B_c^*)+\bar{c}$ and $c+\bar{b}\to B_c(B_c^*)$. For the
hadronic production via $c+\bar{b}\to B_c(B_c^*)$, i.e. the fifth
term in Eq.(\ref{pqcdf0}) when a $p_t$ cut (not too tiny) is put
on (as usually done in experiment), it makes no contributions to
the $B_c(B_c^*)$ production, because its hard subprocess is a
$2\to 1$ process essentially. Thus, we will not consider it in
this paper, because the Tevatron and LHC experiments always put on
some cut for small $p_t$. That is, below we will concentrated
merely on the `intrinsic heavy quark mechanisms' of $B_c(B_c^*)$
production via the two sub-processes: $g+c\to B_c(B_c^*)+b$ and
$g+\bar{b}\to B_c(B_c^*)+\bar{c}$.

\begin{figure}
\centering
\includegraphics[width=0.80\textwidth]{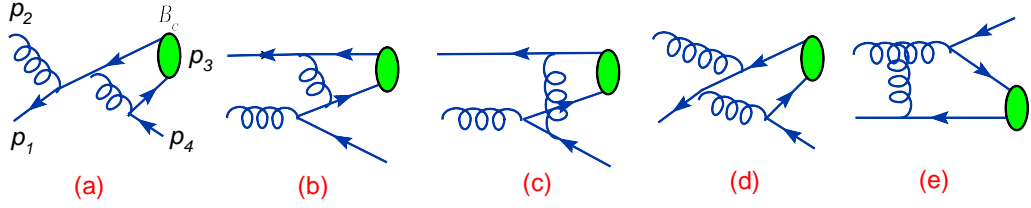}
\caption{Typical Feynman diagrams for bottom induced process:
$g(p_2)+\bar{b}(p_1)\to B_c(p_3)+\bar{c}(p_4)$. The Feynman
diagrams for charm induced process: $g+c\to B_c+b$ can be obtained
by the replacements: $\bar{b}\to c$ and $\bar{c}\to b$.}
\label{fig} \vspace{-0mm}
\end{figure}

\begin{figure}
\centering
\includegraphics[width=0.80\textwidth]{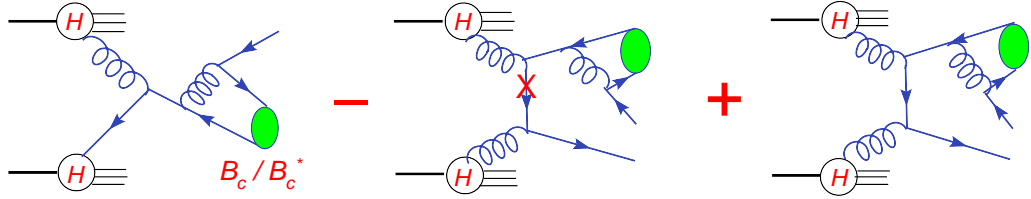}
\caption{Graphical representation for the subtraction method
within GM-VFN \cite{acot}. The subtraction term (diagram) is
placed in the middle to emphasize its similarity both to the
bottom mechanism (left) and the gluon-gluon fusion mechanism
(right). The symbol $\times$ on the internal quark line in the
subtraction term indicates it is close to the mass-shell and
collinear to the gluon and hadron momentum. (As `named' in the
text, the first term and the second one with minus sign are
considered as a whole and called as `intrinsic one'.)}
\label{graphical} \vspace{-0mm}
\end{figure}

Typical Feynman diagrams for these two sub-processes at leading
order (LO) are shown in Fig.(\ref{fig}). The corresponding
diagrams for $B_c^*$ production are similar, and can be obtained
by directly replacing $B_c$ with $B_c^*$ in Fig.(\ref{fig}). To be
specific, according to Eq.(\ref{pqcdf0}), the inclusive $B_c$
hadronic production via `intrinsic charm/bottom mechanisms' can be
formulated as,
\begin{equation}
d\sigma=\sum_{ij}\int dx_{1}\int
dx_{2}F^{i}_{H_{1}}(x_{1},\mu_{F})\times
F^{j}_{H_{2}}(x_{2},\mu)d\hat{\sigma}_{ij\rightarrow
B_{c}(B_c^*)X}(x_{1},x_{2},\mu)\,, \label{pqcdf}
\end{equation}
where $i\neq j$ and $i,j=g,c$ for `intrinsic' charm mechanism and
$i,j=g,\bar{b}$ for `intrinsic' bottom mechanism. Here, the heavy
quark PDF $F^{Q}_{H}(x,\mu)$ ($x=x_1$ or $x_2$, $Q=c$ or
$\bar{b}$, $H=H_1$ or $H_2$), includes the subtraction term
$F^{Q}_{H}(x,\mu)_{SUB}$ as is defined in Eq.(\ref{subtraction})
in order to avoid the double counting. Using subtraction method to
avoid the double counting was pointed out in Ref.\cite{cwz} and
then as the GM-VFN scheme was developed in \cite{gmvfn1,gmvfn2}.
In Fig.(\ref{graphical}), we take the `intrinsic' bottom process
(Fig.(\ref{fig}a)) as an example to illustrate this approach
graphically, which is similar to the case of hadronic production
of heavy quarks \cite{gmvfn1}. The symbol $\times$ on the internal
quark line in the subtraction terms mean that the heavy quark four
momentum squared is on mass-shell and moving longitudinally, which
is a good approximation when the quark is collinear to the gluon,
and results in a factor of order $\alpha_s$ distribution of a
quark in a gluon, like in Eq.(\ref{quark}).

In Eq.(\ref{pqcdf}), $d\hat{\sigma}_{ij\rightarrow
B_{c}X}(x_{1},x_{2},\mu_{F},\mu)=d\hat{\sigma}_{ij\rightarrow
B_{c}X}(x_{1},x_{2},\mu)$ stands for the usual 2-to-2 differential
cross-section,
\begin{equation}
d\hat{\sigma}_{ij\rightarrow
B_{c}(B_c^*)X}(x_{1},x_{2},\mu)=\frac{(2\pi)^4|\overline{M}|^2}
{4\sqrt{(p_1\cdot p_2)^2-m_1^2m_2^2}}
\prod_{i=3}^{4}\frac{d^3\mathbf{p}_i}
{(2\pi)^3(2E_i)}\delta\left(\sum_{i=3}^{4}p_i-p_1-p_2\right)\; .
\end{equation}
Here, $p_1$, $p_2$ are the corresponding momenta of the initial
two partons and $p_3$, $p_4$ are the momenta of the final ones,
respectively. The initial-parton spin and color average and the
final-state quantum number summation are all attributed to the
$|\overline{M}|^2$. According to GM-VFN scheme, the heavy quark
masses are kept in the evaluation of S-matrix. For shortening the
text, we put the explicit expression of amplitude squared,
$|\overline{M}|^2$, in the Appendix instead. We generate the
amplitude and square it $|\overline{M}|^2$ with the program
package: Feynman Diagram Calculation (FDC)\cite{fdc}, which is a
Reduce and Fortran package to perform Feynman diagram calculation
automatically and has been well-tested by various applications.
The phase space integration is manipulated by the routines RAMBOS
\cite{rkw} and VEGAS \cite{gpl}, which can be found in BCVEGPY
\cite{bcvegpy1,bcvegpy2}.

\section{Numerical results and discussions}

In this section, we present the numerical results of the
`intrinsic charm and bottom mechanisms' and then make a comparison
with that of the gluon-gluon fusion mechanism. The combined
results for both the `intrinsic mechanisms' and the gluon-gluon
fusion mechanism are consistently treated within the GM-VFN
scheme. Finally, we shall make a comparison of the present results
with the existing results in the literature.

\subsection{The results for the `intrinsic' charm and bottom mechanisms}

In doing the numerical calculations, the values of the radial wave
function of $B_c$ or $B_c^*$ at the origin is taken as what in
Refs. \cite{chen,potential}, i.e., $|R(0)|^2=1.54 GeV^3$; the
masses of $c$ and $b$ quarks are taken as $m_c=1.50$ GeV and
$m_b=4.80$ GeV, respectively; the mass of the bound state is
approximately taken to be the sum of the two heavy quark masses,
i.e. $M=m_b+m_c=6.30$ GeV (to ensure the gauge invariance of the
concerned amplitudes).

There are a couple of uncertainties remaining in the theoretical
estimations for the $B_c$ meson hadronic production
\cite{changwu}, such as those from the choice of renormalization
scale, factorization scale, {\it etc.}. In present calculations,
the factorization and the renormalization energy scales are set to
be the `transverse mass' of the bound state, i.e. $Q=M_t\equiv
\sqrt{M^2+p_{t}^2}$, where $p_t$ is the transverse momentum of the
bound state; here CTEQ6HQ \cite{6hqcteq} for PDF and the leading
order $\alpha_s$ running above $\Lambda^{(n_f=4)}_{QCD}=0.326$ GeV
are adopted. CTEQ6HQ is adopted here in the calculations, because
it is an improved set of the parton distributions determined for
the GM-VFN scheme that incorporates heavy flavor mass effects.

\begin{figure}
\centering
\includegraphics[width=0.460\textwidth]{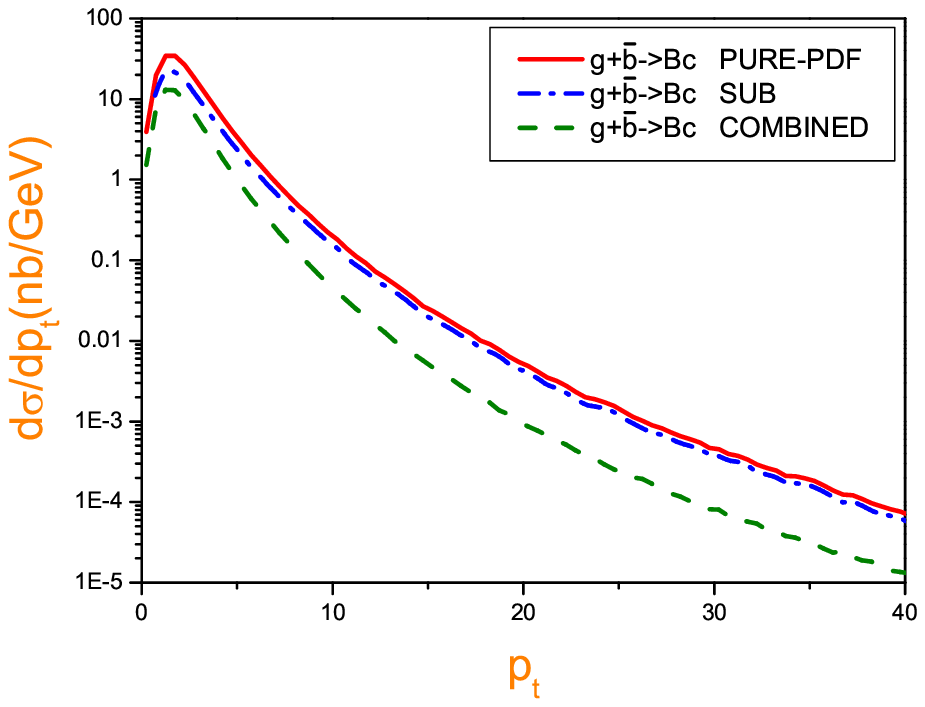}%
\hspace{5mm}
\includegraphics[width=0.460\textwidth]{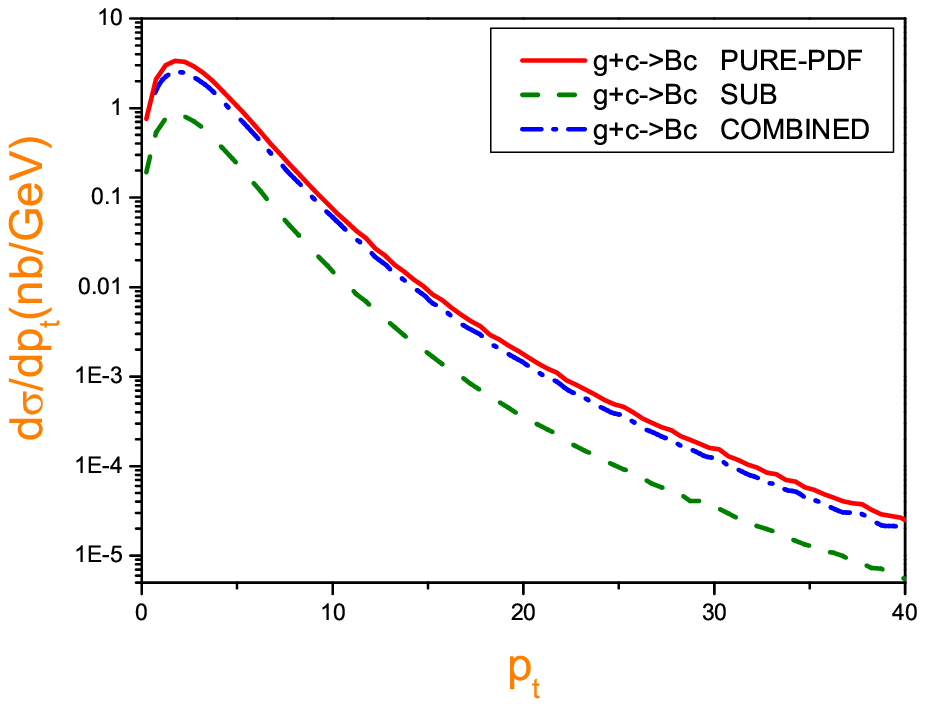}\hspace*{\fill}
\caption{The $p_t$-distributions of the `intrinsic mechanisms' of
the hadronic production of $B_c$ at the LHC. The left diagram is
for $g+\bar{b}\to B_c+\bar{c}$ and the right diagram is for
$g+c\to B_c+b$, where the `PURE-PDF' ones mean the heavy quarks'
PDFs are taken just to be CTEQ6HQ, `SUB' ones mean the heavy
quarks' PDFs are taken just to be the subtraction term defined in
Eq.(\ref{subtraction}), while `COMBINED' ones are the combination
of the `PURE-PDF' and `SUB' components with proper sign as
indicated by Eq.(\ref{pqcdf0}), which correspond to the `intrinsic
charm $g+c\to B_c+b$ or bottom $g+\bar{b}\to B_c+\bar{c}$'
precisely.} \label{fig5} \vspace{-0mm}
\end{figure}

To see the fact how the `double counting' is subtracted in GM-VFN
scheme precisely, first of all we take LHC as an example to
compute each term numerically according to Eq.(\ref{pqcdf0}) for
heavy quarks and present them in Fig.(\ref{fig5}). From the
figure, one may see the consequence of subtraction of the `double
counting' quite substantially. There is large cancellation between
the contributions from the `pure' terms (with heavy quark's PDF
taken to be CTEQ6HQ) and the subtraction terms (with heavy quark's
PDF taken to be the subtraction terms defined in
Eq.(\ref{subtraction})), especially, in the large $p_t$ regions.
Below when we compare the GM-VFN scheme with FFN scheme for the
production, we will, in fact, compare the results for the
production obtained by sum of the gluon-gluon fusion mechanism and
the `intrinsic heavy quark mechanisms' in GM-VFN scheme with those
obtained by gluon-gluon fusion mechanism with PDFs extended to
CTEQ6L in FFN scheme.

\begin{table}
\begin{center}
\caption{The cross-section (in unit of nb) for the hadronic
production of $B_c$ at LHC ($14.0$ TeV) and TEVATRON ($1.96$ TeV),
where for simplicity the symbol $g+\bar{b}$ means $g+\bar{b}\to
B_c+\bar{c}$ and etc. In the calculations, $p_{t}>4GeV$ is taken.
$|y|\leq 1.5$ for LHC, while $|y|\leq 0.6$ at TEVATRON. $q$ stands
for the sum of all the light quarks ($u$, $d$ and $s$).}
\vspace{2mm}
\begin{tabular}{|c||c|c|c|c|c||c|c|}
\hline\hline - & \multicolumn{7}{|c|}{~~~$B_c(^{1}S_{0})$~~~}\\
\hline - & \multicolumn{5}{|c||}{~~~CTEQ6HQ (GM-VFN)~~~}&
\multicolumn{2}{|c|}{~~~CTEQ6L (FFN)~~~}\\
\hline - &~~~$g+\bar{b}$~~~ & ~~~$g+c$ ~~~& ~~~$q+\bar{q}$~~~ &
~~~$g+g$~~~ & ~~~{\it total}~~~& ~~~$q+\bar{q}$~~~ & ~~~$g+g$~~~ \\
\hline\hline LHC & 3.05  & 0.743  & $20.3\times 10^{-3}$
& 6.84 & 10.6 & $17.0\times 10^{-3}$& 12.0\\
\hline TEVATRON & 0.224 & 0.0668 & $4.02\times 10^{-3}$
& 0.414 & 0.709& $3.28\times 10^{-3}$& 0.542\\
\hline\hline
\end{tabular}
\label{cutcross1}
\end{center}
\end{table}

\begin{table}
\begin{center}
\caption{The cross-section (in unit of nb) for the hadronic
production of $B_c^*$ at LHC ($14.0$ TeV) and TEVATRON ($1.96$
TeV), where for simplicity the symbol $g+\bar{b}$ means
$g+\bar{b}\to B_c^*+\bar{c}$ and etc. In the calculations,
$p_{t}>4GeV$ is taken. $|y|\leq 1.5$ for LHC, while $|y|\leq 0.6$
at TEVATRON. $q$ stands for the sum of all the light quarks ($u$,
$d$ and $s$).} \vspace{2mm}
\begin{tabular}{|c||c|c|c|c|c||c|c|}
\hline\hline - & \multicolumn{7}{|c|}{~~~$B_c^*(^{3}S_{1})$~~~}\\
\hline - & \multicolumn{5}{|c||}{~~~CTEQ6HQ (GM-VFN)~~~}&
\multicolumn{2}{|c|}{~~~CTEQ6L (FFN)~~~}\\
\hline - &~~~$g+\bar{b}$~~~ & ~~~$g+c$ ~~~& ~~~$q+\bar{q}$~~~ &
~~~$g+g$~~~ &~~~{\it total}~~~& ~~~$q+\bar{q}$~~~ & ~~~$g+g$~~~ \\
\hline\hline LHC & 11.1  & 4.65  &  $99.1\times 10^{-3}$  & 17.3
&33.1 & $82.8\times 10^{-3}$& 30.7\\
\hline TEVATRON & 0.809 & 0.421 & $19.5\times 10^{-3}$  & 1.03
& 2.28 & $16.0\times 10^{-3}$& 1.34\\
\hline\hline
\end{tabular}
\label{cutcross2}
\end{center}
\end{table}

In high energy hadronic collisions, events with a small $p_t$
and/or a large rapidity $y$ are hard to be detected by detectors,
or say very difficult to be reconstructed among the backgrounds.
So in experimental observations, the events with small $p_t$ and
large rapidity $y$ are dropped practically. Therefore, in
theoretical estimates, proper cuts on $p_t$ and $y$ are applied.
TABLEs \ref{cutcross1} and \ref{cutcross2} show the cross-sections
for the hadronic production of the $B_c(B_c^*)$ meson at LHC and
TEVATRON respectively with cuts: $p_{t}>4$GeV for both LHC and
TEVATRON, and $|y|\leq 1.5$ for LHC, $|y|\leq 0.6$ for TEVATRON.
In TABLEs \ref{cutcross1} and \ref{cutcross2}, we also show the
results from CTEQ6L \cite{6lcteq}, which will be discussed later.
The tables show that the cross-sections of the `intrinsic charm
and bottom mechanisms' are comparable to the gluon-gluon fusion
mechanism, and the sequential order for the cross-sections is
$\sigma_{gg}>\sigma_{g\bar{b}}>\sigma_{gc}>>\sigma_{q\bar{q}}$.
Because the contributions from the light quark and anti-quark
annihilation mechanism are quite small comparing to the other
mechanisms, below we will neglect them.

\begin{figure}
\centering
\includegraphics[width=0.460\textwidth]{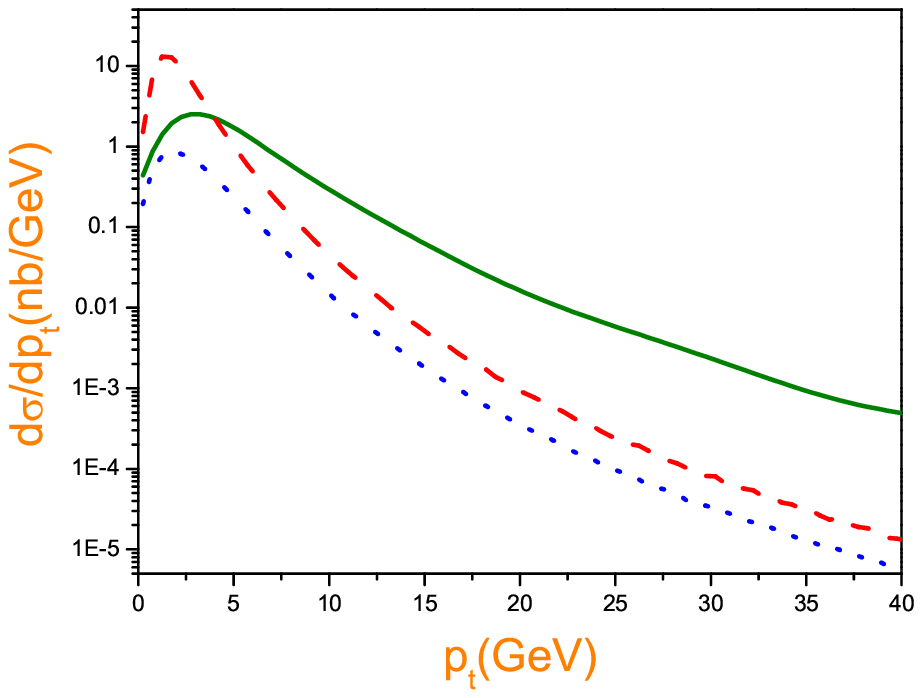}%
\hspace{5mm}
\includegraphics[width=0.460\textwidth]{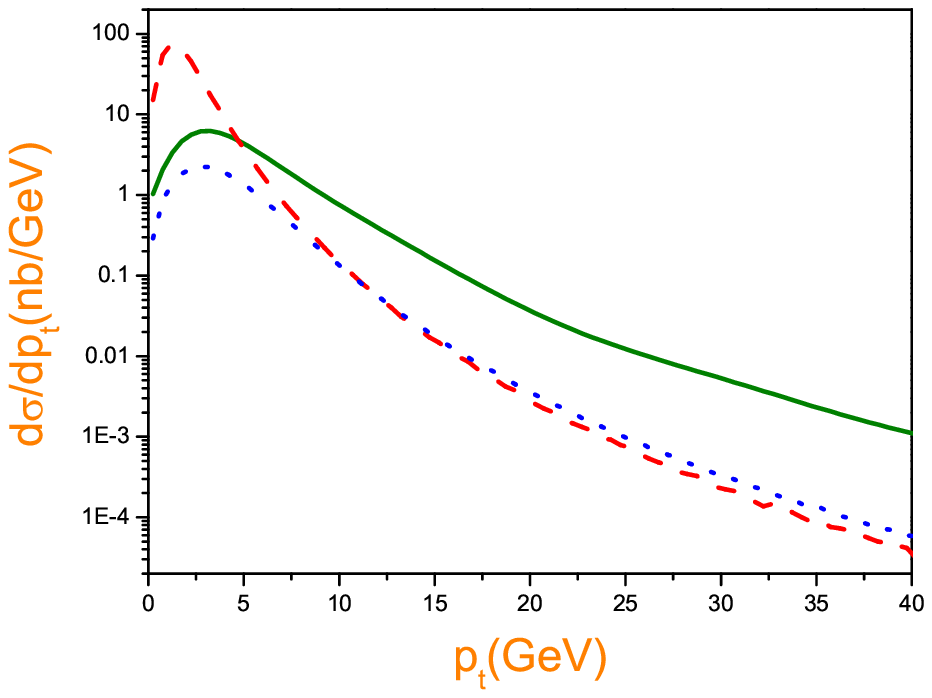}\hspace*{\fill}
\caption{\small The $p_t$ distributions of the hadronic production
of $B_c$ (left diagram) and $B_c^*$ (right diagram) at LHC in
GM-VFN scheme. The solid line is for the $g+g\to B_c(B_c^*)$
mechanism, the dash line is for the intrinsic $g+\bar{b}\to
B_c(B_c^*)$ mechanism and the dotted line is for the intrinsic
$g+c\to B_c(B_c^*)$ mechanism. All the $p_t$ distributions are
drawn with $|y|<1.5$ and the PDF is taken to be CTEQ6HQ.}
\label{fig1} \vspace{-0mm}
\end{figure}

\begin{figure}
\centering
\includegraphics[width=0.460\textwidth]{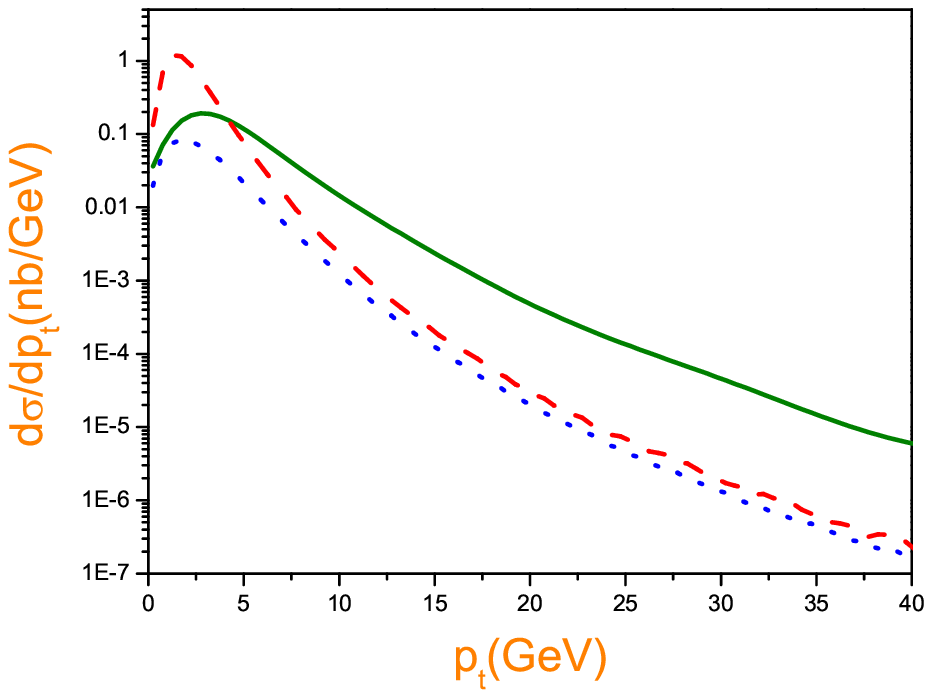}%
\hspace{5mm}
\includegraphics[width=0.460\textwidth]{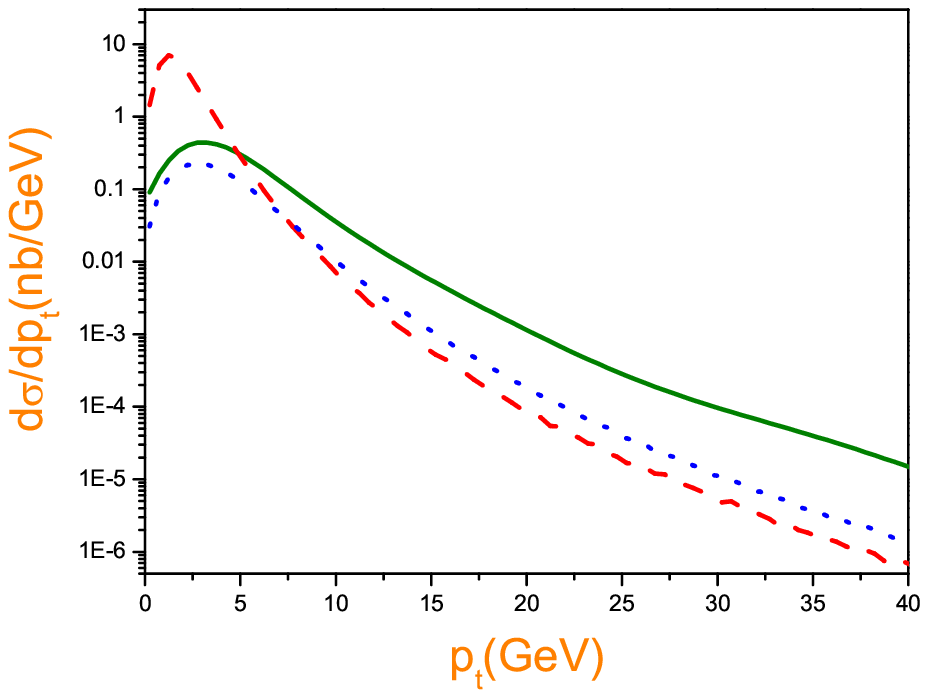}\hspace*{\fill}
\caption{\small The $p_t$ distributions for the hadronic
production of $B_c$ (left diagram) and $B_c^*$ (right diagram) at
TEVATRON in GM-VFN scheme. The solid line is for the $g+g\to
B_c(B_c^*)$ mechanism, the dash line is for the intrinsic
$g+\bar{b}\to B_c(B_c^*)$ mechanism and the dotted line is for the
intrinsic $g+c\to B_c(B_c^*)$ mechanism. All the $p_t$
distributions are drawn with $|y|<0.6$ and the PDF is taken to be
CTEQ6HQ.} \label{fig2} \vspace{-0mm}
\end{figure}

The contributions from the `intrinsic charm and bottom mechanisms'
are remarkable in small $p_t$ region of the $B_c(B_c^*)$
production. To show this point clearly, we present the the
transverse momentum distributions of $B_c(B_c^*)$ in
Figs.(\ref{fig1}, \ref{fig2}). From Figs.(\ref{fig1}, \ref{fig2}),
one may observe that in the small $p_t$ regions, the `intrinsic
charm and bottom mechanisms' are comparable or even greater than
that of the gluon-gluon fusion mechanism. One point worthy to
mention is that even though the `intrinsic charm mechanism' are
smaller than that of `intrinsic bottom mechanism' for the
production in the small $p_t$ region for both $B_c$ and $B_c^*$
production, whereas in the high $p_t$ region, the `intrinsic
charm' contribution becomes slightly larger than that of
`intrinsic bottom' contribution for the case of $B_c^*$. The
contributions from both `intrinsic charm and bottom mechanisms'
drop much more rapidly with the increment of $p_t$ than those from
the gluon-gluon fusion mechanism.

\subsection{Comparisons between the results}

In the literature, only the gluon-gluon fusion mechanism for the
hadronic production of $B_c(B_c^*)$ is considered in most of the
calculations, while the light quark and anti-quark annihilation
mechanism, being less important, is dropped. In these
calculations, which can be considered as an extension of the FFN
scheme as explained at the above, therefore there is no `intrinsic
charm/bottom mechanisms'. In the FFN scheme, the active partons in
the initial state are limited: only light quarks $n_f=3$ and
gluons, while the heavy charm and/or bottom quarks appear only in
the final state. To be consistent with the exact FFN scheme, the
PDFs for the initial partons should be CTEQ5F3 \cite{cteq5}, the
one generated by using the evolution kernels with effective flavor
number $n_{eff}=3$. However, in the references
\cite{prod,prod0,prod1,prod2,prod3,prod4,prod5,prod6}, to do the
production estimates, CTEQ6L for PDFs is used instead of CTEQ5F3.
It is shown in Ref.\cite{changwu} that generally the uncertainties
caused by different leading order PDFs cannot be very great ($\leq
10\%$). In fact, our numerically calculation shows that it only
leads to a small difference $\sim 5\%$ with CTEQ6L to replace
CTEQ5F3 in the estimates.

In TABLEs \ref{cutcross1} and \ref{cutcross2} the cross-sections
obtained with GM-VFN scheme and FFN scheme (with slight extension
CTEQ6L to replace CTEQ5F3) for the hadronic production of
$B_c(B_c^*)$ are shown quantitatively. Due to the fact that the
gluon distribution of CTEQ6HQ for GM-VFN scheme is always smaller
than that of CTEQ6L, especially at small $x$ regions, so the
cross-section for the gluon-gluon fusion under the GM-VFN scheme
is smaller than that under the FFN scheme. Moreover since $x$ may
reach to much smaller region at LHC than at TEVATRON for the
production, the difference between these two schemes is bigger at
LHC than that at TEVATRON. TABLEs \ref{cutcross1}, \ref{cutcross2}
shows this point clearly: at LHC, the cross-section for
gluon-gluon fusion under the GM-VFN is only $\sim 60\%$ of the
case of FFN, while at TEVATRON, such ratio is changed to $\sim
80\%$. When taking the `intrinsic heavy quark mechanisms' into
account for the GM-VFN scheme, one may find that the gap between
the GM-VFN results and the FFN results can be shrunk sizably.
\begin{figure}
\centering
\includegraphics[width=0.460\textwidth]{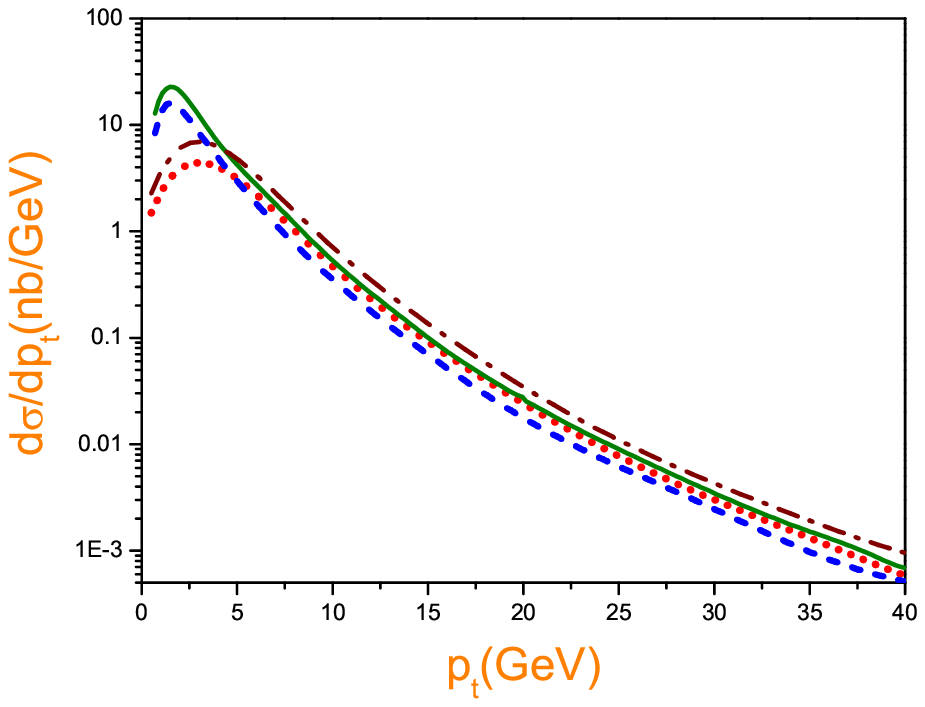}%
\hspace{5mm}
\includegraphics[width=0.460\textwidth]{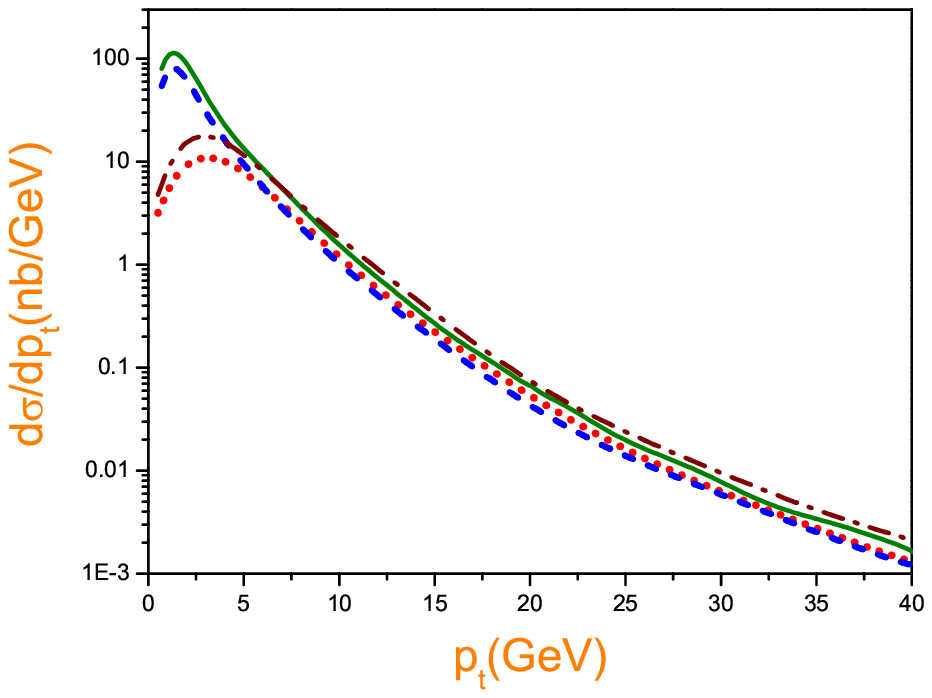}\hspace*{\fill}
\caption{\small  The $p_t$ distributions of the hadronic
production of $B_c$ (left diagram) and $B_c^*$ (right diagram) at
LHC. The solid and the dashed lines stand for the total (sum of
the `intrinsic heavy quark mechanisms' and the gluon-gluon fusion
mechanism) results obtained by the GM-VFN scheme for rapidity cuts
$|y|<2.5$ and $|y|<1.5$ respectively; and the dash-dot and the
dotted lines for gluon-gluon fusion results obtained by the FFN
scheme with PDFs CTEQ6L for rapidity cuts $|y|<2.5$ and $|y|<1.5$
respectively.} \label{fig3} \vspace{-0mm}
\end{figure}

\begin{figure}
\centering
\includegraphics[width=0.460\textwidth]{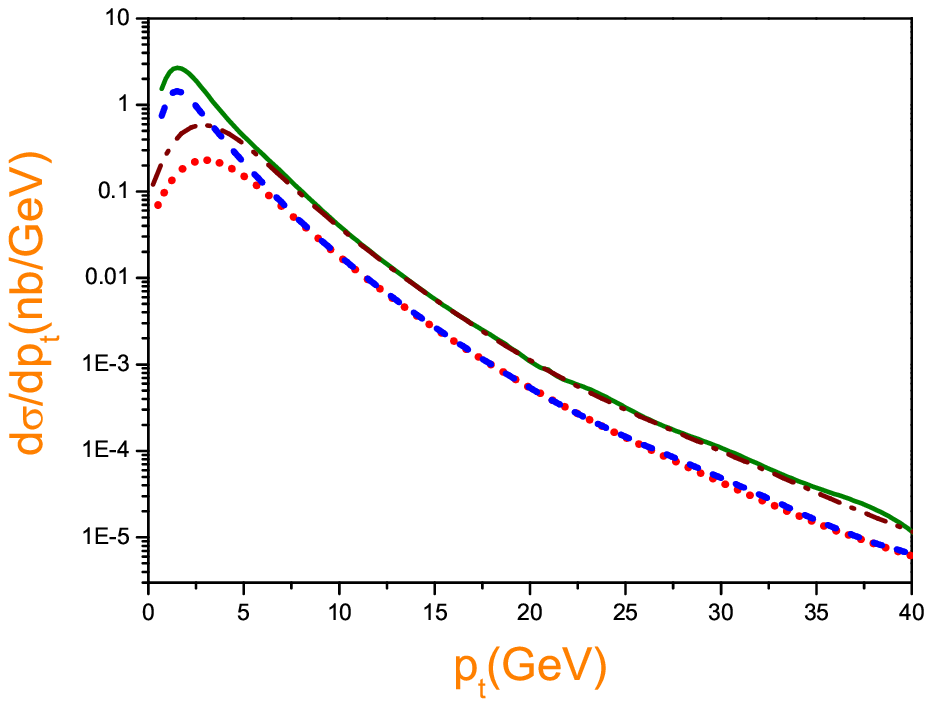}%
\hspace{5mm}
\includegraphics[width=0.460\textwidth]{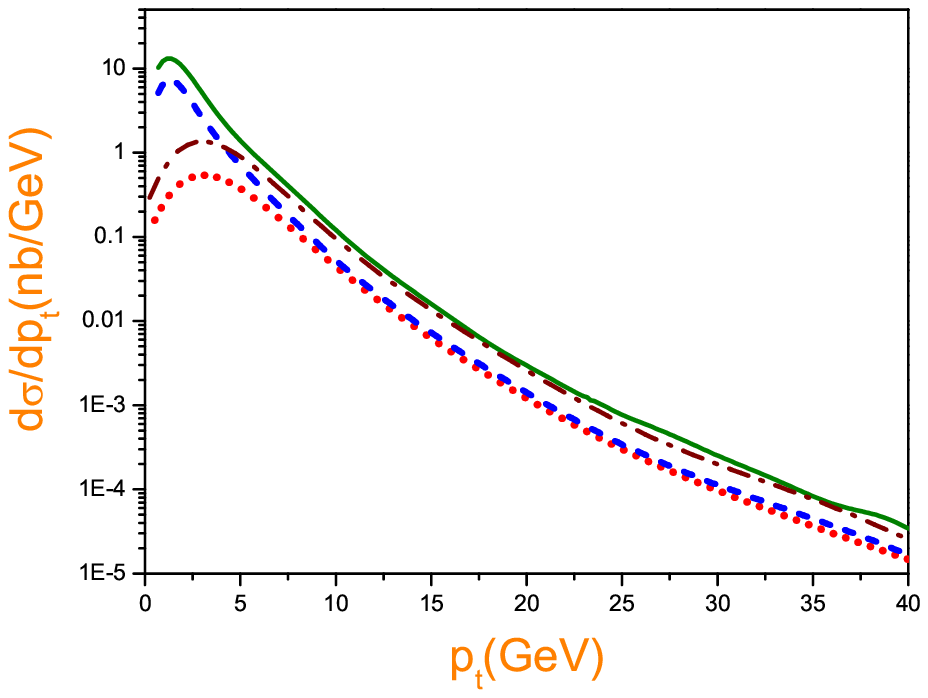}\hspace*{\fill}
\caption{\small  The $p_t$ distributions of the hadronic
production of $B_c$ (left diagram) and $B_c^*$ (right diagram) at
TEVATRON. The solid and the dashed lines stand for the total (sum
of the `intrinsic heavy quark mechanisms' and the gluon-gluon
fusion mechanism) results obtained by the GM-VFN scheme for
rapidity cuts $|y|<1.5$ and $|y|<0.6$ respectively; and the
dash-dot and the dotted lines for gluon-gluon fusion results
obtained by the FFN scheme with PDFs CTEQ6L for rapidity cuts
$|y|<1.5$ and $|y|<0.6$ respectively.} \label{fig4} \vspace{-0mm}
\end{figure}

In fact, the difference of the predictions by the GM-VFN scheme
and the extended FFN scheme is mainly in small $p_t$ region. When
$p_t$ becomes `big', according to the separation of the GM-VFN
scheme, the contributions from the `intrinsic heavy quark
mechanisms' are greatly suppressed due to the cancellation from
the subtraction terms so as to avoid the double counting with
those from the gluon-gluon fusion mechanism, hence the
contributions from the gluon-gluon fusion mechanism become
dominant in the GM-VFN scheme. The sum of the contributions from
the `intrinsic heavy quark mechanisms' and the gluon-gluon fusion
mechanism in the GM-VFN scheme happens to close to those from FFN
scheme, where only the gluon-gluon fusion mechanism alone is
dominant. In order to see the fact precisely, we present the $p_t$
distributions predicted by the two schemes for the hadronic
production of $B_c(B_c^*)$ at LHC and TEVATRON with two possible
rapidity cuts in FIGs.(\ref{fig3}) and (\ref{fig4}) respectively.
From FIGs.(\ref{fig3}) and (\ref{fig4}), one may observe clearly
that the main difference between the predictions by GM-VFN scheme
and the extended FFN scheme (in literatures) is only in small
$p_t$ region ($p_t\lesssim 6.0\sim 7.0$ GeV). In higher $p_t$
regions ($p_t\geq 6.0\sim 7.0$ GeV), the curves predicted by the
two schemes become very close to each other.

\section{discussion and summary}

We have studied the hadronic production of $B_c(B^*_c)$ meson
induced by the heavy quarks inside the incident hadrons under the
GM-VFN scheme. The double counting problem of the gluon-gluon
mechanism with the `intrinsic heavy quark mechanisms' is
consistently solved in the GM-VFN scheme, and also the heavy quark
mass effects, which are important for precise investigations, are
taken into account in both heavy quark PDFs and the hard
scattering kernel.

As shown in TABLEs (\ref{cutcross1}) and (\ref{cutcross2}), we
have found that the total cross sections of the `intrinsic charm
and the bottom mechanisms' may be comparable to that of the
gluon-gluon fusion process, especially the bottom quark
interaction mechanism. The $B_c$ production from the `intrinsic
heavy quarks' drops much more rapidly with the increment of the
transverse momentum $p_t$ than that from the gluon-gluon fusion
mechanism, although it may `overrun' the later in the small
transverse momentum region. At about $6.0\sim 7.0$ GeV the
differential cross section of the $B_c$ transverse production rate
due to the gluon-gluon fusion mechanism surpasses all that from
the involved `intrinsic charm and bottom mechanisms'.

The calculations show clearly that the difference between the
GM-VFN scheme and the FFN scheme (adopted in most of the
calculations in the literature) is mainly around the small $p_t$
region$p_t\lesssim 6.0\sim 7.0$ GeV, as shown in Figs.(\ref{fig3})
and (\ref{fig4}). While the value of $p_t$ becomes big, the
results are almost the same for both schemes. This is because that
when the magnitude of $p_t$ becomes big, the contributions from
the `intrinsic heavy quark mechanisms' are greatly suppressed due
to the subtraction terms in GM-VFN scheme and the dominant
contribution tends to be the gluon-gluon fusion mechanism. For the
most practical purposes and for the production at LHC and Tevatron
one does not need to care of the difference on the predictions of
the GM-VFN scheme and the extended FFN scheme which were used in
most references except those in the small region ($p_t\lesssim
6\sim 7$ GeV).

In conclusion, if one takes the GM-VFN scheme, the `intrinsic
heavy quark mechanisms' in $B_c(B_c^*)$ hadroproduction are
important in small $p_t$ region in comparison with the dominant
gluon-gluon fusion mechanism. Therefore, if $p_t$-cut of $B_c$ can
be taken so small as $4$ GeV, the contributions from the
`intrinsic heavy quark mechanisms' may be seen by the difference
from the predictions not only of the gluon-gluon fusion mechanism
in GM-VFN scheme, but also of the mechanism in extended FFN scheme
(the predictions in most references). Hence a detailed
experimental study of the production in small $p_t$ region could
gain some information on the `intrinsic charm and bottom'
distributions inside the hadron, and will give a `judgement' about
the GM-VFN scheme and the extended FFN scheme. According to the
investigation here, the `intrinsic heavy quark mechanisms' can be
used as a supplement to the usual gluon-gluon fusion mechanism in
GM-VFN scheme, especially, when experiments can reach to very
small $p_t$ region and indicate some deficiency for the
production. Probably some suitable fixed target experiments, in
which the detector may cover almost all solid angles (almost
without $p_t$ cut), can test the `intrinsic charm and bottom
mechanisms' in the future.

\vspace{10mm}

\noindent
{\bf\Large Acknowledgments:}
This work was supported in part by the Natural Science Foundation
of China (NSFC). C.H. Chang would like to thank W.K. Tung for
private communication on the parton distribution function CTEQ.\\

\noindent {\large\bf Appendix: The squared amplitudes for the
sub-processes}

For convenience, we express the square of the amplitudes by the
Mandelstam variants $s$, $t$ and $u$, which are defined as:
\begin{displaymath}
s=(p_{1}+p_{2})^2,~~~t=(p_{1}-p_{3})^2,~~~u=(p_{1}-p_{4})^2 ,
\end{displaymath}
where $p_i=(E_i,p_{ix},p_{iy},p_{iz})$ are the corresponding
momenta for the involved particles: $p_1$ and $p_2$ are the
momenta of initial partons, $p_3$ and $p_4$ are the momenta of
$B_c(B_c^*)$ and another outgoing particles respectively. Further
more, for $\bar{b}(p_1)+g(p_2)\to
B_c(p_3)/B_c^*(p_3)+\bar{c}(p_4)$, we set
\begin{displaymath}
u_{1}=(u-(m_b+m_c)^2),~~~s_{1}=(s-m_{b}^2),~~~t_{1}= (t-m_{c}^2).
\end{displaymath}
The relation, $u_1+t_1+s_1=0$, is useful to make all the
expressions for the square of the amplitudes compact.

The square of the amplitude for the subprocess
$\bar{b}(p_1)+g(p_2)\to B_c(p_3)+\bar{c}(p_4)$ can be written as,
\begin{eqnarray}
|\overline{M}|^2 &&={{\alpha_s}^3{f_{B_c}}^2{\pi}^4 \over 2^3 3^6
M m_{b}^2m_{c}^2}\Bigg(m_{c}^6({{-64s_{1} \over t_{1}^3}}+{{- 16
\over t_{1}^2}}+{{127 \over s_{1}t_{1}}})+16m_{c}^8( {{-8s_{1}
\over t_{1}^3u_{1}}}+{{-9 \over t_{1}^2u_{1}
}})+2m_{c}^5m_{b}({{-64s_{1} \over t_{1}^3}}+  \nonumber \\
&&{{-96 \over t_{1}^2}}+{{109 \over s_{1}t_{1}}}) +8m_{c}^7
m_{b}({{-80s_{1} \over t_{1}^3u_{1}}}+{{-148 \over
t_{1}^2u_{1}}}+{{-61 \over s_{1}t_{1}u_{1}}})+8m_{c}^9m_{b}(
{{64s_{1}^2 \over t_{1}^4u_{1}^2}}+{{144s_{1} \over
t_{1}^3u_{1}^2}}+{{81 \over t_{1}^2u_{1}^2}})+
\nonumber \\
&&32 m_{c}^2m_{b}^2( {{-u_{1} \over t_{1}^2}}+{{-u_{1} \over
s_{1}^2}})+m_{c}^4m_{b}^2({{-608 \over t_{1}^2}}+{{- 527 \over
s_{1}t_{1}}}+{{-272 \over s_{1}^2}}+{{64 t_{1} \over
s_{1}^3}})+16m_{c}^6m_{b}^2({{-112s_{1} \over t_{1}
^3u_{1}}}+\nonumber \\
&&{{-335 \over t_{1}^2u_{1}}}+{{-287 \over s_{1}t_{1}u_{1}}}+{{-65
\over s_{1}^2u_{1}}})+16m_{c}^8 m_{b}^2({{128s_{1}^2 \over
t_{1}^4u_{1}^2}}+{{296 s_{1} \over t_{1}^3u_{1}^2}}+{{163 \over
t_{1}^2u_{1}^2}}+ {{-9 \over s_{1}t_{1}u_{1}^2}})+   \nonumber \\
&&2 m_{c}^3m_{b}^3({{ 107 \over s_{1}t_{1}}}+{{-304u_{1}^2 \over
s_{1}^2t_{1}^2}}+ {{32u_{1}^4 \over s_{1}^3t_{1}^3}})+8m_{c}^5
m_{b}^3( {{-352s_{1} \over t_{1}^3u_{1}}}+{{-1432 \over t_{1}^2
u_{1}}}+{{-1747 \over s_{1}t_{1}u_{1}}}+{{-724 \over
s_{1}^2u_{1}}}+{{-80t_{1} \over s_{1}^3u_{1}}})  \nonumber \\
&&+8m_{c}^7m_{b}^3 ({{384s_{1}^2 \over t_{1}^4u_{1}^2}}+{{928s_{1}
\over t_{1}^3u_{1}^2}}+{{351 \over t_{1}^2u_{1}^2}} + {{-364 \over
s_{1}t_{1}u_{1}^2}}+{{-143 \over s_{1}^2 u_{1}^2}})+8
m_{c}^4m_{b}^4({{-20 \over s_{1}t_{1}u_{1}}}+{{-199u_{1} \over
s_{1}^2t_{1}^2}}+{{-136u_{1}^3 \over
s_{1}^3t_{1}^3}})   \nonumber \\
&& +16 m_{c}^6m_{b}^4( {{128s_{1}^2 \over
t_{1}^4u_{1}^2}}+{{336s_{1} \over t_{1}^3u_{1}^2}}+{{-118 \over
t_{1}^2u_{1}^2}}+{{-647 \over s_{1}t_{1}u_{1}^2}}+{{-285 \over
s_{1}^2u_{1}^2}}+ {{8t_{1} \over s_{1}^3u_{1}^2}})+4
m_{c}^5m_{b}^5( {{-70 \over s_{1}t_{1}u_{1}^2}} \nonumber \\
&&+{{-1025 \over s_{1}^2t_{1}^2}}+{{-176u_{1}^2 \over
s_{1}^3t_{1}^3}}+ {{64u_{1}^4 \over
s_{1}^4t_{1}^4}})+m_{b}^6({{127 \over s_{1}t_{1}}}+{{-16 \over
s_{1}^2}}+{{-64t_{1}\over s_{1}^3}})+64({{-m_{b}^4u_{1} \over
t_{1}^2}}+ {{-2m_{b}^3m_{c}u_{1} \over
t_{1}^2}})\Bigg)\nonumber\\
& & +\Bigg(s_{1}\leftrightarrow t_{1},~~~m_b\leftrightarrow
m_c\Bigg),
\end{eqnarray}
where $(s_{1}\leftrightarrow t_{1},~~~m_b\leftrightarrow m_c)$
stands for the remaining terms that can be directly obtained by
exchanging $s_{1}\leftrightarrow t_{1}$ and
$m_b\leftrightarrow m_c$ for all the terms in the first big
parenthesis. The square of the amplitude for the subprocess
$c(p_1)+g(p_2)\to B_c(p_3)+b(p_4)$ can be directly obtained from
the above formula by taking the transition $m_b\leftrightarrow
m_c$.

The square of the amplitude for the subprocess
$\bar{b}(p_1)+g(p_2)\to B_c^*(p_3)+\bar{c}(p_4)$,
\begin{eqnarray}
|\overline{M}|^2 &&={{{\alpha_s}^3{f_{B_c}}^2{\pi}^4 \over 2^3 3^6
m_{b}^2m_{c}^ 2 M}}\Bigg(64m_{c}^4({{-u_{1} \over s_{1}^2}}+{{-2
u_{1}^3 \over s_{1}^2t_{1}^2}})+m_{c}^6({{64s_{1}^3 \over t_{1}^3
u_{1}^2}}+{{624s_{1}^2 \over t_{1}^2u_{1}^2}}+{{1469 s_{1} \over
t_{1}u_{1}^2}}+{{1294 \over u_{1}^2}}+{{
381t_{1} \over s_{1}u_{1}^2}}) \nonumber  \\
&&+48m_{c}^8({{-8s_{1} \over t_{1}^3u_{1}}}+{{-9 \over
t_{1}^2u_{1}}})+128m_{c}^3m_{b}( {{-u_{1} \over
s_{1}^2}}+{{-4u_{1}^3 \over s_{1}^2
t_{1}^2}})+2m_{c}^5m_{b}({{192s_{1}^3 \over t_{1}^3u_{1}^2}}+
{{1840s_{1}^2 \over t_{1}^2u_{1}^2}}+{{4393s_{1}
\over t_{1}u_{1}^2}} \nonumber  \\
&&+{{3982 \over u_{1}^2}}+ {{1225t_{1} \over
s_{1}u_{1}^2}})+8m_{c}^7m_{b}({{ -496s_{1} \over
t_{1}^3u_{1}}}+{{-732 \over t_{1}^2u_{1}}}+ {{-183 \over
s_{1}t_{1}u_{1}}})+24m_{c}^9m_{b}({{ 64s_{1}^2 \over
t_{1}^4u_{1}^2}}+{{144s_{1} \over
t_{1}^3u_{1}^2}}+ {{81 \over t_{1}^2u_{1}^2}}) \nonumber  \\
&&+32m_{c}^2m_{b}^2( {{2u_{1} \over s_{1}t_{1}}}+{{-13u_{1}^3
\over s_{1}^2 t_{1}^2}})+m_{c}^4m_{b}^2({{768s_{1}^3 \over
t_{1}^3u_{1}^2}}+ {{7424s_{1}^2 \over
t_{1}^2u_{1}^2}}+{{19363s_{1} \over t_{1}u_{1}^2}}+{{20978 \over
u_{1}^2}}+{{9955 t_{1} \over s_{1}u_{1}^2}}  \nonumber \\
&&+{{1936t_{1}^2 \over s_{1}^2u_{1} ^2}}+{{192t_{1}^3 \over
s_{1}^3u_{1}^2}})+16m_{c}^6m_{b}^2( {{-720s_{1} \over
t_{1}^3u_{1}}}+{{-1325 \over t_{1}^2 u_{1}}}+{{-701 \over
s_{1}t_{1}u_{1}}}+{{-195 \over
s_{1}^2u_{1}}})+48m_{c}^8m_{b}^2({{128s_{1}^2 \over t_{1}^4
u_{1}^2}}  \nonumber \\
&&+{{296s_{1} \over t_{1}^3u_{1}^2}}+{{ 163 \over
t_{1}^2u_{1}^2}}+{{-9 \over
s_{1}t_{1}u_{1}^2}})+2m_{c}^3m_{b}^3({{20 \over u_{1}^2}}+
{{119 \over s_{1}t_{1}}}+{{616u_{1}^2 \over s_{1}^2 t_{1}^2}}
+{{160u_{1}^4 \over s_{1}^3t_{1}^3}})+
8m_{c}^5 m_{b}^3({{-1824s_{1} \over t_{1}^3u_{1}}}\nonumber   \\
&&+{{-4488 \over t_{1}^2u_{1}}}+{{-3961 \over s_{1}t_{1}u_{1}}}+
{{-1948 \over s_{1}^2u_{1}}}+{{-240t_{1} \over s_{1}^3
u_{1}}})+24m_{c}^7m_{b}^3({{384s_{1}^2 \over t_{1}^4u_{1}^2}}
+{{928s_{1} \over t_{1}^3u_{1}^2}}+{{351 \over
t_{1}^2u_{1}^2}}+{{-364 \over s_{1}t_{1}u_{1}^2}}\nonumber   \\
&&+ {{-143 \over s_{1}^2u_{1}^2}})+8 m_{c}^4m_{b}^4({{260 \over
s_{1}t_{1}u_{1}}}+{{107u_{1} \over s_{1}^2t_{1}^2}}+{{ -536u_{1}^3
\over s_{1}^3t_{1}^3}})+48m_{c}^6m_{b}^4({{ 128s_{1}^2 \over
t_{1}^4u_{1}^2}}+{{336s_{1} \over t_{1}^3u_{1}^2}} +{{-118 \over
t_{1}^2u_{1}^2}} \nonumber  \\
&&+{{-647 \over s_{1}t_{1} u_{1}^2}}+{{-285 \over
s_{1}^2u_{1}^2}}+{{8t_{1} \over s_{1}^3u_{1}^2}})+12 m_{c}
^5m_{b}^5({{-70 \over s_{1}t_{1}u_{1}^2}}+{{-1025 \over
s_{1}^2t_{1}^2}}+{{-176u_{1}^2 \over s_{1}^3t_{1}^3}}+ {{64u_{1}^4
\over s_{1}^4t_{1}^4}})\Bigg)\nonumber\\
& & +\Bigg(s_{1}\leftrightarrow t_{1},~~~m_b\leftrightarrow
m_c\Bigg),
\end{eqnarray}
where $(s_{1}\leftrightarrow t_{1},~~~m_b\leftrightarrow m_c)$
stands for the remaining terms that can be directly obtained by
exchanging $s_{1}\leftrightarrow t_{1}$ and
$m_b\leftrightarrow m_c$ for all the terms in the first big
parenthesis. The square of the amplitude for the subprocess
$c(p_1)+g(p_2)\to B_c^*(p_3)+b(p_4)$ can be directly obtained from
the above formula by taking the transition $m_b\leftrightarrow
m_c$.

\end{document}